\documentclass{emulateapj}
\usepackage{apjfonts}
\newcommand{\Mmin}{M_{\mathrm{min}}}

\shortauthors{CONROY ET AL.}
\shorttitle{The Varied Fates of $\lowercase{z}\sim2$ Star-Forming Galaxies}

\begin{document}

\title{The Varied Fates of $\lowercase{z}\sim2$ Star-forming Galaxies}

\author{
Charlie Conroy\altaffilmark{1},
Alice E. Shapley\altaffilmark{1,2,3},
Jeremy L. Tinker\altaffilmark{4}, 
Michael R. Santos\altaffilmark{5,6},
Gerard Lemson\altaffilmark{7,8}
}


\altaffiltext{1}{Department of Astrophysical Sciences, Princeton
  University, Princeton, NJ 08544}
\altaffiltext{2}{Alfred P. Sloan Fellow}
\altaffiltext{3}{David and Lucile Packard Fellow}
\altaffiltext{4}{Kavli Institute of Cosmological Physics, University
  of Chicago, Chicago, IL 60637}
\altaffiltext{5}{Space Telescope Science Institute, 3700 San Martin
  Dr., Baltimore, MD 21218}
\altaffiltext{6}{Giacconi Fellow}
\altaffiltext{7}{Astronomisches Rechen-Institut, Zentrum f\"ur
  Astronomie der Universit\"at Heidelberg, Moenchhofstr. 12-14, 69120
  Heidelberg, Germany}
\altaffiltext{8}{Max-Planck Institut f\"ur extraterrestrische Physik,
  Giessenbach Str., 85748 Garching, Germany}

\begin{abstract}

  Star-forming galaxies constitute the majority of galaxies with
  stellar masses $\gtrsim10^{10}\,h^{-2} M_\Sun$ at $z\sim2$.  It is
  thus critical to understand their origins, evolution, and connection
  to the underlying dark matter distribution.  To this end, we
  identify the dark matter halos (including subhalos) that are likely
  to contain star-forming galaxies at $z\sim2$ (z2SFGs) within a large
  dissipationless cosmological simulation and then use halo merger
  histories to follow the evolution of z2SFG descendants to $z\sim1$
  and $z\sim0$.  The evolved halos at these epochs are then confronted
  with an array of observational data in order to uncover the likely
  descendants of z2SFGs.  Though the evolved halos have clustering
  strengths comparable to red galaxies at $z\sim1$ and $z\sim0$, we
  find that the bulk of z2SFGs {\it do not} evolve into red galaxies,
  at either epoch. This conclusion is based primarily on the fact that
  the space density of z2SFGs is much higher than that of lower
  redshift red galaxies, even when accounting for the merging of z2SFG
  descendants, which decreases the number density of z2SFG descendants
  by at most a factor of two by $z\sim0$.  Of the $\sim50$\% of z2SFGs
  that survive to $z\sim0$, $\sim70$\% reside at the center of
  $z\sim0$ dark matter halos with $M>10^{12} \,h^{-1} M_\Sun$.  Halo
  occupation modeling of $z\sim0$ galaxies suggests that such halos
  are occupied by galaxies with $M_r\lesssim-20.5$, implying that
  these z2SFGs evolve into ``typical'' $\sim L^\ast$ galaxies today,
  including our own Galaxy.  The remaining $\sim30$\% become satellite
  galaxies by $z\sim0$, and comparison to halo occupation modeling
  suggests that they are rather faint, with $M_r\lesssim-19.5$.  These
  conclusions are at least a partial departure from previous work due
  primarily to the increased accuracy of observational data at
  $z\lesssim 1$, and to higher resolution $N$-body simulations that
  explicitly follow the evolution of dark matter subhalos, whose
  observational counterparts are likely satellite galaxies.  These
  conclusions are qualitatively generic in the sense that any halo
  mass-selected sample of galaxies at one epoch will evolve into a
  more complex and heterogeneous sample of galaxies at a later epoch.
  This heterogeneity is driven largely by the fact that some galaxies
  will continue to accrete matter and form stars throughout their
  evolution, while others will become satellites and thus have their
  growth suppressed relative to galaxies in the field.

\end{abstract}


\section{Introduction}
\label{s:intro}

The past decade has witnessed an enormous increase in our knowledge of
the high-redshift ($z>2$) Universe.  The unique rest-frame UV spectral
signatures of star-forming galaxies have allowed observers to select
galaxies at $z\gtrsim2$ based solely on optical photometry with very
high efficiency \citep[e.g.][]{Steidel03, Steidel04}.  Extensive
multi-wavelength follow-up of these objects, including optical and
near-IR spectroscopy, and imaging at wavelengths from X-ray through
radio, has yielded estimates of rest-frame UV luminosities and colors
\citep{Steidel99, Adelberger00, Reddy07}, stellar masses
\citep{Shapley05, Erb06b}, star-formation rates \citep{Erb06a,
  Reddy06a}, chemical abundances \citep{Pettini01, Erb06c}, and
clustering strengths \citep{Adelberger03, Adelberger05, Giavalisco01,
  Ouchi05, Lee06, Hamana06, Quadri07} of a large, well-defined
population of galaxies that dominate the star-formation rate density
\citep{Reddy07} at an epoch when the Universe was only a few billion
years old. Complementary techniques tuned to the rest-frame optical
properties of galaxies at similar epochs have identified objects with
typically redder colors and higher mass-to-light (M/L) ratios, but the
corresponding spectroscopic confirmation has only been obtained for
the brightest, rare objects \citep{Franx03, Daddi04, vanDokkum06}.

At the same time, models connecting galaxies to the underlying dark
matter distribution have been put forth which either focus on
quantifying the statistical relation between galaxies and dark matter
halos \citep[e.g.][]{Berlind02, Bullock02, Yan03, Seljak00,
  Scoccimarro01, Zheng04, Kravtsov04, Conroy06a}, or predicting the
connection based on physical principles \citep[e.g.][]{White91,
  Somerville99, Cole00, Hatton03, Springel01, Croton06,
  Bower06}. Essential to these latter efforts has been the vast
increase in size and resolution of $N$-body simulations, which has
only recently allowed for the construction of detailed merger trees
that follow not only the evolution of distinct halos, i.e. those halos
not contained within any larger halo, but also subhalos --- the halos
contained within the virial radii of distinct halos --- whose
observational counterparts are likely satellite galaxies
\citep[e.g.][]{Springel01}.  Throughout we refer to both distinct
halos and subhalos generically as halos.  Despite impressive advances
in our understanding both of the properties of halos themselves and of
the relation between galaxies and halos, fundamental questions remain.

The origin and fate of z2SFGs is one such question.  Herein we define
z2SFGs as those galaxies selected to lie at $z\sim2$ based on optical
$U_nG\mathcal{R}$ photometry that are brighter than
$\mathcal{R}=25.5$, which corresponds to a rest-frame UV magnitude
limit at this epoch.  Historically, UV-selected high-redshift galaxies
have been described as either low-mass, merger-induced starbursts or
more massive galaxies ``quiescently'' forming stars.  With early data,
it was not possible to distinguish between these opposing ideas
\citep{Lowenthal97, Coles98, Mo99, Giavalisco01, Kolatt99,
  Somerville01, Wechsler01}.  However, recent modeling in conjunction
with more recent data on the abundance and spatial clustering seems to
favor the latter picture \citep{Conroy06a, Adelberger05}. The
distribution of stellar populations of these objects lends support to
the massive, quiescent scenario as well \citep{Shapley01, Erb06b}.

Once identified at high redshift, it is also important to determine
what type(s) of galaxies z2SFGs evolve into at later epochs.  Most
previous analyses have focused on the so-called Lyman-Break Galaxies,
which are UV-selected star-forming galaxies at $z\sim3$.  We argue in
later sections that these galaxies are qualitatively similar to z2SFGs
and thus here we make no distinction between the two. Some analyses
focusing on the strong observed clustering of z2SFGs favored the
scenario where z2SFGs evolved into the observed massive red galaxies
at $z\sim1$, and by $z\sim0$ were at the centers of rich groups and
clusters \citep{Mo96b, Baugh98, Governato98, Governato01, Blaizot04,
  Adelberger05}, though other models incorporating the high observed
comoving number density of z2SFGs suggested a more nuanced history
\citep{Moustakas02}.  Subsequent to these efforts, there has been a
vast increase in our understanding of the number densities and
clustering of galaxies as a function of luminosity and color both at
$z\sim1$ \citep[e.g.][]{Coil06b, Pollo06, Meneux06, Coil07} and
$z\sim0$ \citep[e.g.][]{Zehavi05}. Better constraints on the
statistics of galaxy populations at lower redshift are crucial for a
robust identification of the descendants of high-redshift
galaxies. These recent data at $z\leq 1$, in conjunction with new
high-resolution simulations and dark-matter-halo merger trees,
motivate a re-evaluation of the nature and subsequent evolution of
z2SFGs.

In the present work we focus on the fate(s) of observed z2SFGs.  This
redshift range is our focus here rather than $z\geq 3$, as there has
been much recent attention to this lower redshift window with several
complementary galaxy selection techniques. Plus, a more extensive set
of multi-wavelength imaging and spectroscopy exists for $z\sim 2$
galaxies, allowing the more robust estimate of a number of physical
properties such as stellar, gas, and dynamical masses, and
extinction-corrected star-formation rates
\citep{Erb06b,Shapley05,ForsterSchreiber06, Reddy06a}, which are
important to consider along with the inferred host dark-matter halo
properties. Such a detailed comparison of halo and galaxy masses and
formation histories is currently not possible at higher redshifts.

Under the assumption of a tight correlation between galaxy light and
halo mass, the observed clustering of z2SFGs puts a constraint on the
minimum halo mass hosting such galaxies ($\S$\ref{s:p}).  Halo merger
trees extracted from a cosmological $N$-body simulation are then used
to follow the halos hosting z2SFGs to later epochs.  Comparing the
clustering, satellite fraction, and number density of these evolved
halos to observed galaxy populations at both $z\sim1$ and $z\sim0$
then provides constraints on the fates of z2SFGs ($\S$\ref{s:r}).

Throughout we assume a $\Lambda$CDM cosmology with
$(\Omega_m,\Omega_{\Lambda},\sigma_8) = (0.25,0.75,0.9)$, consistent
with the {\it WMAP1} data \citep{Spergel03}.  Where applicable, we
leave results in terms of $h$, the Hubble parameter in units of $100$
km s$^{-1}$ Mpc$^{-1}$.  In several sections we discuss how our
results are affected by adopting the cosmological parameters favored
by the {\it WMAP3} data \citep{Spergel07}, where the primary
difference is a lower $\sigma_8$. Halo masses are measured as the mass
interior to a region with mean enclosed density equal to $200$ times
the critical density.  A \citet{Chabrier03} IMF is assumed when
quoting stellar masses.  All magnitudes are quoted in the AB system;
we omit the factor of $5{\rm log} (h)$ when quoting absolute
magnitudes for brevity.

\section{Preliminaries}
\label{s:p}

\subsection{Connecting Galaxies to Halos}
\label{s:conn}

The clustering strength of a given sample of galaxies can be used to
estimate the minimum dark matter halo mass, $\Mmin$, hosting such
galaxies, for a specified $\Lambda$CDM cosmology \citep[see
e.g.][]{Wechsler98, Wechsler01, Adelberger03, Quadri07, Gawiser07}.
This approach assumes that a sample of galaxies above a given
luminosity threshold corresponds, at least approximately, to a sample
of halos above a given mass threshold.  Under this assumption, the
threshold $\Mmin$ is varied until the clustering of halos with
$M\geq\Mmin$ matches that of the observational sample in question.

While early attempts only considered distinct halos, recent work has
demonstrated that this approach can be refined by including dark
matter subhalos --- halos that orbit within the potential wells of
distinct halos --- as possible sites for galaxies
\citep[e.g.][]{Colin99, Kravtsov99b, Springel01, Kravtsov04,
  Tasitsiomi04, Vale04, Vale06, Conroy06a}.  The approach is as simple
as before --- $\Mmin$ is varied until a match with the observed
correlation function is obtained --- except now galaxies can reside
within subhalos, and hence there can be multiple galaxies per distinct
halo.  This refinement is desirable because the observed correlation
function requires there to be multiple galaxies per distinct halo
\citep[e.g.][]{Zehavi05}.

In this model, the space density of halos with $M\geq\Mmin$ need not
match the space density of observed galaxies.  This approach thus
allows the possibility of there being fewer than one galaxy per halo
or subhalo, which, if favored by the data, would indicate that z2SFGs
are ``on'' only a fraction of the time \citep{Martini01}.  Such a
scenario may arise if the star-formation in these galaxies is episodic
due to, for example, major mergers.  The converse possibility, where
there is more than one galaxy per halo, would signal an inconsistency
in the model, since the model assumes that halos (both distinct and
subhalos) can host at most one galaxy.

Previous work has demonstrated that better agreement with observed
small-scale clustering data can be achieved when using the subhalo
mass at the time when it is accreted on to a larger halo (at the
``epoch of accretion''), rather than its present day mass
\citep{Nagai05, Vale06, Conroy06a, Wang06, Berrier06}.  The accretion
epoch mass is expected to correlate more strongly with galaxy
properties because, while a subhalo can experience mass-loss due to
tidal stripping, the galaxy, which is embedded at the very center of
the halo, is much less affected by tidal evolution.  For subhalos, we
use this accretion-epoch mass herein when counting halos above
$\Mmin$, but note that the dynamical times at $z\sim2$ are short
($\sim0.1$ Gyr) and thus subhalos merge rapidly. Therefore, those that
are identified at $z\sim2$ were only accreted rather recently.  In
fact, more than half of the subhalos identified in the simulation at
$z\sim2$ above our best-fit $\Mmin$ (see below) were accreted within
the last simulation output, which at these epochs is $\sim0.2$ Gyr.
Because of their recent accretion, they have thus lost little mass due
to the effects of tidal stripping; the average halo has lost $30$\% of
its mass since accretion, as inferred from a high-resolution $N$-body
simulation (see below).  The difference between using accretion-epoch
versus current masses should thus be small; we include it only for
completeness.

In sum, we vary $\Mmin$ until the clustering of the halos with
$M\geq\Mmin$ matches the clustering of the observed z2SFGs.  We
include both distinct halos and subhalos so that there can be multiple
galaxies per distinct halo (as there are multiple galaxies per
group/cluster).  For subhalos we use the subhalo mass at the epoch of
accretion as this should be a better proxy for the stellar, and hence
luminous, content of the subhalo.

As applied to z2SFGs, the approach outlined above assumes that the
rest-frame UV luminosity of galaxies is tightly and monotonically
correlated with dark matter halo mass.  Since this does not appear to
be the case at $z\sim0$ \citep[e.g.][]{Heinis07}, such an assumption
may at first glance appear unjustified.  In fact, however, several
lines of evidence suggest that this assumption is valid at $z\sim2$.
The most straightforward evidence comes from the fact that the large
scale clustering strength of observed z2SFGs is an increasing function
of rest-frame UV luminosity \citep{Adelberger05} at the
$\sim2\sigma$ level.  Larger samples at somewhat higher redshifts
($3<z<5$) confirm this trend \citep{Ouchi05, Lee06}.  Such facts are
most readily understood if there is a correlation between UV
luminosity and halo mass since higher mass halos are more strongly
clustered than lower mass halos.  At lower redshift, this general line
of reasoning is routinely employed \citep[e.g.][]{Zehavi05,
  Conroy06a}, and has been verified by more direct measurements of
halo mass, such as weak gravitational lensing
\citep[e.g.][]{Mandelbaum06}.

Further evidence comes from the observation of a clear correlation
between the star-formation rate and baryonic mass (stellar mass and
cold gas mass within $\sim 6$ kpc) for z2SFGs \citep{Erb06a, Erb06b}.
If the vast majority of galaxies at $z\sim2$ are forming stars
vigorously enough to be detected as z2SFGs, as appears to be the case
\citep{Franx03, Daddi04, vanDokkum06}, then a selection on rest-frame
UV luminosity, which traces star-formation -- modulo the effects of
dust extinction -- should correspond to a selection on baryonic mass.
Since a strong redshift-independent correlation between baryonic and
halo mass is a firm prediction of any theory of galaxy formation
\citep[e.g.][]{Crain07}, a UV luminosity-limited sample is
approximately a halo mass-selected sample at these epochs.

There is thus ample evidence that rest-frame UV luminosity is tightly
correlated with halo mass at $z\sim2$.  However, the lack of any
strong dependence of clustering on UV luminosity at $z\sim0$ suggests
that this is not the case in the local Universe
\citep[e.g.][]{Heinis07}.  We now briefly provide a plausible
explanation for this difference between high and low redshift.

At high redshift, star-formation and baryonic mass are monotonically
related primarily because most of the gas in halos at high redshift is
cold ($\sim 10^4 $K), and so it directly contributes to the
star-formation rate \citep[e.g.][]{Keres05}.  Only galaxies within the
most massive and hence rarest halos have had star-formation largely
truncated at this epoch \citep{Franx03, Daddi04, vanDokkum06,
  Quadri07}.  At later times, infalling gas in massive halos is
shock-heated to the virial temperature of the halo \citep[$\sim 10^7
$K;][]{White78, Keres05, Dekel06, Cattaneo07}.  High temperature gas
is much more susceptible to further heating processes, and thus we can
assume that once gas is shock-heated, it remains hot forever
\citep[observations indicate that this must be the case for the hot
gas that permeates groups and clusters of galaxies;][]{Peterson03}.
The transition from predominantly low- to high-temperature gas
quenches star-formation in high mass halos, thereby breaking the
monotonic relation between star formation and baryonic (and thus halo)
mass.  These trends are generically reproduced in hydrodynamic
cosmological simulations \citep[e.g.][]{Blanton00, Keres05,
  Cattaneo07}, and provide a justification of our assumption that UV
light and halo mass are tightly correlated at high redshift, despite
the absence of such a trend at lower redshift.

\subsection{Data at $\lowercase{z}\sim2$}

The observed z2SFGs are selected by $U_nG\mathcal{R}$ color cuts and
required to be brighter than $\mathcal{R}=25.5$.  These are objects
satisfying the ``BX/MD'' criteria of \citet{Steidel04, Steidel03}, but
their comoving number density has been corrected for the
incompleteness arising from star-forming galaxies scattering out of
the $U_nG\mathcal{R}$ color selection region due to photometric errors
\citep[see e.g.][]{Adelberger04, Adelberger05, Reddy07}.  This sample
is thus meant to encompass all star-forming galaxies at $z\sim2$ with
$\mathcal{R}\leq 25.5$.  As estimated from the rest-frame UV
luminosity function, the number density of these galaxies is $11\times
10^{-3}\,h^{3}$ Mpc$^{-3}$, with an uncertainty of $\sim10$\%
\citep{Reddy07}.  This is $\sim80$\% higher than the preliminary
number density estimates reported in \citet{Adelberger05}.  

The photometric limit in the observed $\mathcal{R}$-band corresponds
to the rest-frame UV at these epochs.  At $z\sim2$ $M^\ast(1700
${\AA}$)=-20.2$ and $\mathcal{R}=25.5$ corresponds to $M^\ast(1700
${\AA}$)=-18.6$; the z2SFG sample thus extends $\sim1.6$ magnitudes
below $M^\ast$ \citep{Reddy07}.  This rest-frame UV limit
approximately corresponds to an unobscured star-formation rate
threshold of $15\,M_\Sun$ yr$^{-1}$, given the typical z2SFG
rest-frame UV extinction factors of $\sim 4-5$ \citep{Steidel04,
  Reddy07}.

The angular clustering of these z2SFGs has been measured by
\citet{Adelberger05}, who find that the observed clustering can be fit
with a real-space (de-projected) correlation function that is a
power-law, $\xi=(r/r_0)^{-\gamma}$, with $\gamma=1.6\pm0.1$ and
$r_0=4.2\pm0.5\,h^{-1}$ Mpc\footnote{The de-projection of the observed
  angular correlation function that yields estimates of $r_0$ and
  $\gamma$ is cosmology-dependent.  The results reported by
  \citet{Adelberger05} assumed a cosmology where
  $(\Omega_m,\Omega_{\Lambda},\sigma_8) = (0.30,0.70,0.9)$ ---
  slightly different from the one adopted herein.  We expect that the
  change in $r_0$ induced by updating the cosmology is within
  $1\sigma$ of the value reported above.}.  Note that while the number
density has been completeness corrected, the clustering measurements
have not.  That is, \citet{Adelberger05} only measured the correlation
function for objects satisfying the $U_nG\mathcal{R}$ color criteria.
We are thus assuming that the objects not included in the clustering
analysis have the same clustering properties as those that were
included.  Since $\mathcal{R}\leq 25.5$ star-forming galaxies were
missed from the color selection due primarily to photometric scatter
(i.e. by random processes), this assumption is well motivated
\citep{Reddy07}.

\subsection{$N$-Body Simulations}
\label{s:sim}

\begin{figure}[t]
\plotone{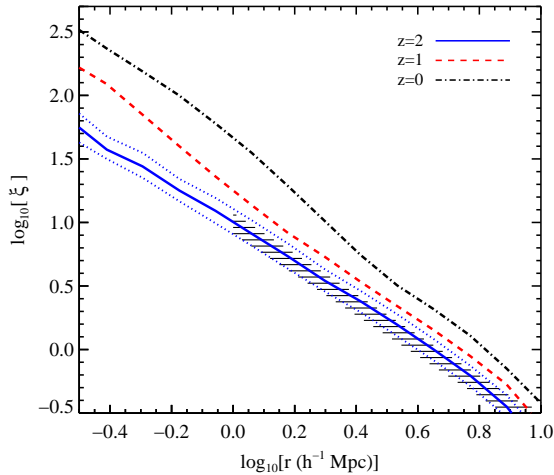}
\vspace{0.5cm}
\caption{The correlation function for halos with $M\geq10^{11.4}
  h^{-1} M_\Sun$ at $z\sim2$ ({\it solid line}) that have clustering
  properties similar to observed z2SFGs \citep[{\it hatched
    region};][]{Adelberger05}.  Also plotted are the correlation
  functions of those halos evolved to $z\sim1$ and $z\sim0$.  The
  correlation function increases toward lower redshift both because of
  the increased clustering of halos and because of the higher
  satellite fraction at later times, which increases the weight given
  to higher mass (more clustered) halos.  Halos with masses
  $M\geq10^{11.2} h^{-1} M_\Sun$ and $M\geq10^{11.6} h^{-1} M_\Sun$ at
  $z\sim2$ are included for comparison ({\it lower and upper dotted
    lines, respectively}). }
\vspace{0.5cm}
\label{f:xiev}
\end{figure}

The properties and histories of dark matter halos are taken from the
Millennium simulation, a large cosmological $N$-body simulation with
sufficient resolution to identify both distinct halos and subhalos
\citep{Springel05}.  Here and throughout distinct halos are those
halos that are not contained within any larger virialized system while
subhalos are halos whose centers are contained within the virial radii
of larger systems. This classification is analogous to the distinction
between central and satellite galaxies.  The simulation box has length
$500 \,h^{-1}$ Mpc, particle mass $m_p=8.6\times10^8\,h^{-1} M_\Sun$,
and was run with the following cosmological parameters:
$\sigma_8=0.9$, $\Omega_m = 1-\Omega_\Lambda=0.25$, $h=0.73$, and
$n_s=1$ where $n_s$ is the slope of the initial linear power spectrum.
Halo merger trees, which connect halos across time, also exist for
this simulation\footnote{The merger trees and halo catalogs are
  publically available and can be found here:
  \texttt{http://www.mpa-garching.mpg.de/millennium}.}
\citep{Springel05}, and are utilized below.  In particular, the merger
trees are used to define the accretion-epoch mass for subhalos as the
mass when the subhalo was last considered a distinct halo.  Working
with a simulation of this size can be computationally expensive; for
this reason we thus make use of a $150^3\,h^{-3}$ Mpc$^3$ region of
the simulation.  This size is sufficient for our purposes; for
example, making use of a $250^3\,h^{-3}$ Mpc$^3$ region at $z\sim2$
has a negligible effect on our conclusions at that epoch.

In order to investigate the sensitivity of our results to cosmological
parameters, we performed two additional $N$-body simulations.  One
used the same cosmological parameters as the Millennium run, and the
other used cosmological parameters from the $WMAP3$ results
\citep[$\Omega_m = 0.239$, $\Omega_\Lambda = 0.761$, $\Omega_b =
0.04166$, $h = 0.73$, $n_s = 0.953$, $\sigma_8 =
0.756$;][]{Spergel07}.  Each simulation contained $512^3$ particles in
a box of length $141.3$ $h^{-1}$ Mpc, resulting in a particle mass of
$1.46\times 10^9\,h^{-1}\, M_\sun$ for the Millennium cosmology and
$1.39\times 10^9\,h^{-1}\, M_\sun$ for the $WMAP3$ cosmology. The
GADGET-2 $N$-body code \citep{Springel01a, Springel05} with a
Plummer-equivalent softening length of $7.9\,h^{-1}\,kpc$ was used for
both simulations.

In these additional simulations, we identified distinct halos with a
friends-of-friends algorithm\footnote{The algorithm we use has been
  made freely available by the University of Washington HPCC group:
  \texttt{http://www-hpcc.astro.washington.edu/tools/fof.html}.}
using a linking length of $0.2$ in units of the mean interparticle
separation.  The halos of interest contained at least $70$ particles.
No subhalo detection was performed on these simulations.  We define
halo mass as the mass within a sphere centered on the minimum
potential energy halo particle (after removing unbound particles) and
enclosing a density equal to $200$ times the critical density.  This
definition closely matches the definition of $M_{\rm crit,200}$ used
in the Millennium simulation (which first removed subhalo particles
before identifying the halo particle with minimum potential energy)
and elsewhere in this paper.

The descendant of a $z\sim2$ halo was identified by finding the halo
at a later time that contained the plurality of the particles
comprising the $z\sim2$ halo.  However, the position of the descendant
was taken to be the later-time center-of-mass of the particles
comprising the $z\sim2$ halo.  We explicitly verified that our
simulation with the Millennium cosmology produced consistent distinct
halo mass functions, clustering as a function of halo mass, and halo
descendant clustering as a function of halo mass as the Millennium
simulation.

\section{Results}
\label{s:r}

\subsection{The \lowercase{z}2SFG-Halo Connection}
\label{s:z2}

The clustering of halos extracted from the simulation at $z\sim2$ is
shown in Figure \ref{f:xiev} for our best-fit $\Mmin$ threshold, along
with the observed z2SFG clustering.  As discussed in $\S$\ref{s:sim},
here and throughout ``halos'' refers to both distinct halos and
subhalos.  The clustering of halos with $\Mmin$ $\pm0.2$ dex from the
best-fit $\Mmin$ is included for comparison.  It is clear from the
figure that $\Mmin=10^{11.4} \,h^{-1}\,M_\Sun$ provides a good match
to the observed clustering of z2SFGs.  The slope of the halo
correlation function on scales $1<r<10 \,h^{-1}$ Mpc is $-1.5$ which
is consistent at the $1\sigma$ level with the slope inferred from
observations ($-1.6$).  This figure also contains the clustering of
the $z\sim2$ halos evolved to $z\sim1$ and $z\sim0$, based on the
Millennium simulation merger trees.  The space density of $z\sim2$
halos with $M\geq\Mmin$ is $7.5\times 10^{-3} \,h^{3}$ Mpc$^{-3}$.

The uncertainties of the z2SFG clustering data translates into an
uncertainty on $\Mmin$ of $\sim\pm0.2$ dex (as determined by eye; see
Figure \ref{f:xiev}).  This uncertainty affects both the inferred
number densities and the clustering strength of the corresponding host
halos.  For $\Mmin=10^{11.2} \,h^{-1}\,M_\Sun$ the number density is
$13\times10^{-3}\, h^{3}$ Mpc$^{-3}$ while for $\Mmin=10^{11.6}
\,h^{-1}\,M_\Sun$ it is $4.3\times10^{-3}\,h^{3}$ Mpc$^{-3}$.  This
range brackets the observed number density of z2SFGs ($11\times
10^{-3} \,h^{3}$ Mpc$^{-3}$) and indicates that, within the
uncertainty, every halo above $\Mmin$ contains one z2SFG.  The
uncertainty in $\Mmin$ is explicitly incorporated into our uncertainty
in the clustering of z2SFG halos and their descendants in the
following sections.

In reality the dark matter halo occupation function of z2SFGs need not
be a step function at $\Mmin$ (zero galaxies per halo/subhalo below
$\Mmin$ and one above it).  For example, scatter between galaxy UV
luminosity and halo mass at $z\sim2$ will result in a more gradual
rise of the occupation function from zero to one around $\Mmin$.  In
order to understand the qualitative effect of scatter on our inferred
number density of halos that match the clustering of z2SFGs, we have
run a series of halo occupation models that were constrained to match
the observed correlation function \citep[see e.g.][for
details]{Tinker06}.  In these models both $\Mmin$ and the amount of
scatter between mass and light were left as free parameters.  The
resulting number density of halos with clustering matching the
observed z2SFGs varies by $\sim20$\% compared to the number density of
halos when setting the scatter to zero (which is our default model
herein).  The effect of scatter is thus insignificant for our
purposes.

Using $r_0$ to constrain the minimum dark matter halo mass of
high-redshift galaxies is not a new technique
\citep[e.g.][]{Wechsler98}.  Previously, \citet{Adelberger05} used the
GIF-$\Lambda$CDM simulation \citep{Kauffmann99a} to constrain the
minimum mass, finding that $\Mmin\sim 10^{11.8} \,h^{-1}\, M_\Sun$
provided a good fit to the $z\sim2$ data (which is the same data used
herein).  The difference between this value and ours ($10^{11.4}
\,h^{-1}\, M_\Sun$) is due to the updated $\Omega_m$ and more
physically motivated transfer function in the simulations we
use.\footnote{The GIF simulations used an analytic fitting function
  for the power spectrum transfer function \citep{Bond84} while the
  Millennium simulation utilized the more accurate CMBFAST code
  \citep{Seljak96} to generate it.}  At first glance this may be
worrisome, since the cosmology used herein is already somewhat
out-dated insofar as the new {\it WMAP3} results \citep{Spergel07}
favor a lower normalization of the power spectrum ($\sigma_8=0.76$)
than what we assume.  To explore the dependence of our conclusions on
cosmological parameters, we have made use of simulations, described in
$\S$\ref{s:sim}, with the updated {\it WMAP3} parameters and find that
the best-fit $\Mmin$ decreases by $0.3-0.4$ dex.  This lower $\Mmin$
results in a $35-70$\% increase in the abundance of z2SFG host halos
--- well within the uncertainties associated with matching halos to
galaxies using clustering, as discussed above.  The differences
between the {\it WMAP1} and {\it WMAP3} cosmological parameters thus
have a small impact on the halo-z2SFG connection.

In sum, the observed clustering strength and comoving number density
of z2SFGs combine to suggest that every dark matter halo and subhalo
with mass $\geq10^{11.4}\,h^{-1}\,M_\Sun$ is host to roughly one z2SFG.
The subhalo fraction among these halos implies that 11\% of z2SFGs are
satellites.  

Techniques complementary to the optical color-selection method used to
identify z2SFGs have identified a population of massive, red galaxies
at $z\sim2$ with low space density \citep[i.e. objects with $n\sim
10^{-4} \, h^{3}$ Mpc$^{-3}$;][]{Franx03, Daddi04, vanDokkum06}.  A
significant fraction of these objects may be missed by the z2SFG
$UG\mathcal{R}$ criteria --- especially if they have little or no
ongoing star-formation.  However, since the observed z2SFG sample has
a space density of $n\sim 10^{-2} \, h^{3}$ Mpc$^{-3}$, our results
are not sensitive to this possibly distinct population of massive red
galaxies because such objects are comparatively rare and thus have
little impact on the number density and clustering of the overall set
of z2SFGs.

\subsubsection{Quiescent versus Collisional z2SFGs}

In the previous section we found that the number density of halos
above $\Mmin$ was consistent with the observed number density of
z2SFGs, given the uncertainty in $\Mmin$.  Thus, within the
uncertainties, all halos above $\Mmin$ contain one z2SFG.  Such a
scenario arises naturally if z2SFGs are quiescently forming stars,
since in this picture every sufficiently massive halo should have the
same conditions necessary for star-formation (e.g. a sufficient supply
of cold gas).  Here and throughout, ``quiescent'' refers to an
approximately continuous star-formation history, as opposed to an
episodic one.

However, it is less clear that a collisional starburst scenario (where
star formation is merger-induced and thus episodic) would be
consistent with this result, since the majority of halos, at any mass
scale, are not undergoing violent mergers frequently enough.  For
example, based on merger trees in the Millennium simulation, only
$\sim20$\% of halos above $\Mmin=10^{11.4}\,h^{-1}\, M_\Sun$ at
$z\sim2$ have had a merger with mass ratio $<3:1$ in the past
gigayear.  If elevated star formation occurs for such mergers but not
for mergers at larger mass ratios, as suggested by controlled
hydrodynamic simulations \citep[e.g.][]{Mihos94, Cox07}, then there
are simply too few such mergers in halos with $M\geq\Mmin$ to explain
the abundance of observed z2SFGs under the collisional starburst
scenario.  Previous modeling efforts \citep{Kolatt99, Wechsler01}
concluded that the collisional starburst scenario was viable because
they included collisions with very high mass ratios ($10:1$ and
greater).  It now appears, however, that such merger ratios will not
result in the enhanced star-formation \citep{Cox07} necessary to make
them detected as z2SFGs.  Moreover, the small scatter in the observed
star-formation rate-stellar mass relation at $z\sim1$ suggests that by
this epoch starbursts are not the dominant mode of star formation in
most galaxies \citep{Noeske07a}.

Of course, to accomodate the low rate of roughly equal mass mergers,
one could lower $\Mmin$ until the number density of halos with low
mass-ratio mergers ($<3:1$) matched that of the z2SFGs.  However,
lowering $\Mmin$ in order to incorporate more colliding halos would
also lower the large scale clustering stength, and would quickly
produce disagreement with the observed clustering of z2SFGs
\citep[note that any ``merger bias'', in the sense that colliding
halos are more clustered than other halos of the same mass, is
probably small; see][]{Scannapieco03}.  Models that rely on collisions
as the impetus for star-formation would thus have a very difficult
time simultaneously matching the observed clustering and space density
of observed z2SFGs.

Some degree of caution is in order when attempting to discriminate
between these two scenarios for the origin of z2SFGs, as the accretion
and merging rate of halos is significantly higher at high redshift
compared to $z\sim0$, and thus the dichotomy between quiescent and
collisional star-formation may not be applicable at high redshift.  As
noted above, major mergers, which are the types of mergers thought to
fuel collisional starbursts, are still sufficiently rare to rule that
channel out as a major contributor to the star-formation rate.  But
minor mergers of ratio $\sim10:1$ are far more common at higher
redshift, and far fewer halos are truly quiescent, at least in terms
of their accretion of dark matter.  For example, the halos above
$\Mmin$ at $z\sim2$ in the Millennium simulation have a mass doubling
time of $\sim1$ Gyr; only the rarest and most massive halos have such
vigorous accretion properties today \citep{Wechsler02}.

\subsubsection{The Baryon Budget}

Here we briefly compare the available baryon reservoir inferred from
our results to estimates of the gas+stellar mass in a subsample of
observed z2SFGs.

The average halo mass for the z2SFGs is $10^{11.8} \,h^{-1}\, M_\Sun$;
this can be converted into an average {\it baryonic} mass by assuming
that each halo contains the universal baryon fraction, $f_b=0.17$
\citep{Spergel07}.  Under this assumption, the average baryonic mass
for the z2SFG sample is $1.1\times10^{11} \, h^{-1}\,M_\Sun$.  For a
fraction of the z2SFGs modeled in the present work there exist stellar
and gas mass estimates \citep{Erb06b}.  The observed gas masses were
estimated by utilizing the empirical \citet{Kennicutt98} law that
relates star-formation rate densities to cold gas densities.  The
star-formation rate was estimated from the H$\alpha$ emission line and
its surface density was estimated from the $\sim 6$ kpc extent of the
emission.  Stellar masses were estimated from stellar population
synthesis modeling with a \citet{Chabrier03} IMF.  The average observed
gas+stellar masses of these objects is $5.8\times 10^{10}\, M_\Sun$,
for $h=0.7$.

In order to compare our expected average baryon fraction to this
gas+stellar mass estimate we convert our mass to $h=0.7$ units, and
find that the observed gas+stellar mass estimate is roughly a factor
of three smaller than the average baryonic mass inferred from the
halos.  Based on the assumption that the Kennicutt law holds at
$z\sim2$, and thus that the estimated gas masses are accurate, our
results indicate that on average a third of the baryons in the halo
are in the central $\sim 6$ kpc.  Furthermore, while $>50$\% of the
\emph{observed} baryons in z2SFGs appear to have been converted into
stars \citep{Erb06b,Erb06c}, the inferred baryonic mass from the
average z2SFG halo mass indicates that only $\sim20$\% of the
\emph{total available} baryons have been converted into stars.
Moreover, halos are constantly being fed new gas as they accrete
gas-filled dark matter halos.  There is thus a sufficient supply of
gas to sustain star formation for many gigayears, if the gas is able
to cool.  As we will see below, this feature is important because many
of the z2SFG descendants are galaxies that are still forming stars at
$z\sim1$ and $z\sim0$.

\subsection{\lowercase{z}2SFG Evolution to $\lowercase{z}\sim1$ and
  $\lowercase{z}\sim0$}
\label{s:z10}

The identification of the halos likely containing z2SFGs is the first
step toward understanding the evolution of these galaxies to $z\sim1$
and $z\sim0$, where there are well-defined samples such as the Deep
Extragalactic Evolutionary Probe 2 \citep[DEEP2;][]{Davis03}, the
VIMOS VLT Deep Survey \citep[VVDS;][]{LeFevre05b}, the 2dF Galaxy
Redshift Survey \citep[2dFGRS;][]{Colless01}, and the Sloan Digital Sky
Survey \citep[SDSS;][]{DR4}.  The next step is to evolve these halos
forward in time using the halo merger trees from the Millennium
simulation.  Note that the merger trees follow the evolution of
subhalos until they have been destroyed, i.e., until the halo finder
no longer identifies the subhalo as a bound clump of particles.

\begin{figure}[t]
\plotone{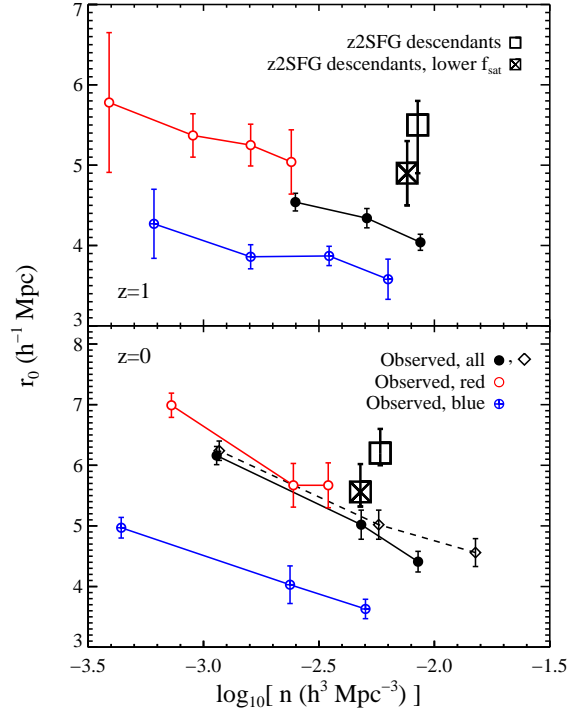}
\vspace{0.5cm}
\caption{Relationship between clustering strength ($r_0$) and sample
  number density ($n$) for observed galaxies and the descendants of
  the halos hosting z2SFGs at $z\sim1$ (\emph{top panel}) and $z\sim0$
  (\emph{bottom panel}).  The data at $z\sim1$ are for samples defined
  above various magnitude thresholds \citep[from $-19.5$ to $-20.5$ in
  half magnitude steps for the overall sample, and from $-19$ to $-21$
  for the color-defined samples;][]{Coil06b, Coil07}, while at
  $z\sim0$ they are defined for magnitude bins \citep[in one magnitude
  intervals from $-19$ to $-22$][]{Zehavi05}.  We have also included
  data from the ``all'' sample at $z\sim0$ for magnitude threshold
  samples ({\it diamonds}) in order to show the differences between
  binned and threshold samples (see $\S$\ref{s:nn} for details).  For
  the halos, error bars encompass the uncertainty due both to $\Mmin$
  and the merger vs. no merger scenarios (see $\S$\ref{s:z10} for
  details).  The location of the symbols for the halos indicates the
  values for the merger scenario with $\Mmin=10^{11.4} \,h^{-1}\,
  M_\Sun$ (\emph{large open boxes}).  z2SFG descendant halos with a
  satellite fraction lowered to match observations are also included
  (\emph{large crossed boxes}).}
\vspace{0.5cm}
\label{f:nvsr0}
\end{figure}

The primary ambiguity in this next step is how we treat z2SFG subhalos
that, according to the merger trees, have merged/disrupted at some
later time.  These cases are ambiguous because the Millennium
simulation may, like any simulation, {\it artificially} destroy a
subhalo, either because of resolution limits \citep[e.g.][]{Moore99,
  Klypin99} or because the simulation does not include the effects of
baryon condensation, which may make a subhalo more resilient to
disruption in the real universe \citep[although the latter effect is
expected to be small for the regimes of interest here; see
e.g.][]{Nagai05, Weinberg06}.  We incorporate this uncertainty by
computing results in this section for two cases that should bracket
the range of possibilities.  In the first case we assume that the
simulation is correct and that when a subhalo merges, so does the
galaxy embedded within it.  In the second case we assume that none of
the subhalos actually merges.  In this case, if a subhalo merges
within a distinct halo according to the merger tree, we place the
satellite within the distinct halo with a position specified by an NFW
distribution \citep{NFW97} appropriate for the background dark
matter\footnote{In fact, the galaxies associated with disrupted
  subhalos will likely be more centrally concentrated than the dark
  matter \citep{Sales07}.  However, the radial distribution of
  galaxies within the halo has a negligible effect on the large-scale
  clustering strength and leaves the satellite fraction and number
  density of the galaxies unchanged.  This uncertainty thus does not
  impact our analysis.}.  These will be referred to as the merger and
no-merger scenarios below.  For reference, the conclusions reached in
\citet{Adelberger05}, namely that z2SFGs evolve into massive red
galaxies by $z\sim0$, were based on the no-merger scenario.

In what follows we focus on three constraints that will help
discriminate between possible evolutionary histories of z2SFGs; these
are the space density of galaxies, their large scale ($1\lesssim
r\lesssim10 \,h^{-1}$ Mpc) clustering strength, and the fraction of
galaxies that are satellites.  These three constraints, when combined,
strongly disfavor any scenario where z2SFGs evolve into a single class
of objects at lower redshift (i.e. red, blue, central satellite).  As
we discuss the evolution of z2SFGs to lower redshifts, it is worth
remembering that z2SFGs are defined according to a luminosity-limit in
the rest-frame UV, while samples at lower redshifts are defined
according to increasingly redder rest-frame bands ($B$-band at
$z\sim1$ and $r$-band at $z\sim0$).  At increasingly shorter
wavelengths, the light emitted by galaxies is increasingly dominated
by young stars and hence recent episodes of star-formation, while
longer wavelengths are dominated by older stars and hence probe the
total stellar mass of a galaxy. At first glance then, connecting
galaxies selected by star-formation at $z\sim2$ to galaxies selected
more closely by stellar mass at $z\sim0$ would seem to be a daunting
task.  However, with the assignment of z2SFGs to dark matter halos,
connecting these galaxies to their lower redshift counterparts becomes
simpler thanks to the halo merger trees, which provide a clear
connection between high and low redshift.

\begin{figure}[t]
\plotone{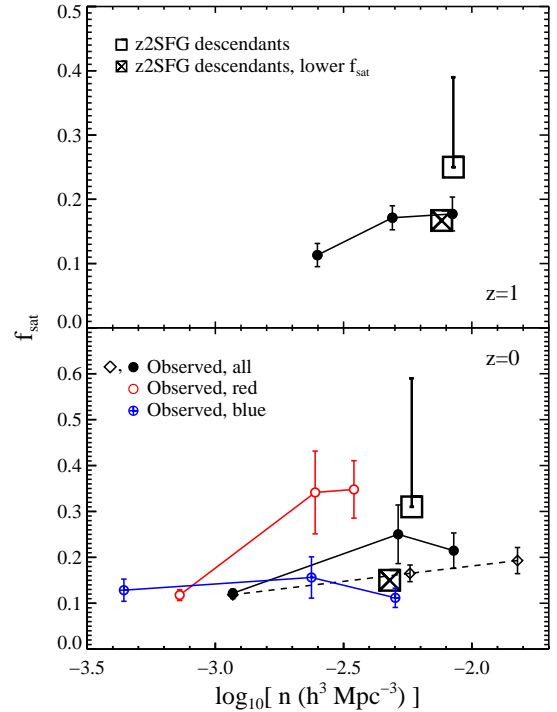}
\vspace{0.5cm}
\caption{Relationship between satellite fraction and sample number
  density for the z2SFG halo descendants ({\it large boxes}) and
  various observations ({\it circles}; symbols are as in Figure
  \ref{f:nvsr0}), at both $z\sim1$ ({\it top panel}) and $z\sim0$
  ({\it bottom panel}).  Also included are z2SFG descendant halos with
  a satellite fraction measured from the Millennium simulation
  (\emph{large open boxes}) and lowered to match observations
  (\emph{large crossed boxes}).  For the halos, error bars encompass
  the uncertainty due both to $\Mmin$ and the merger vs. no merger
  scenarios (see $\S$\ref{s:z10} for details).  The location of the
  symbols for the halos indicates the values for the merger scenario
  with $\Mmin=10^{11.4} \,h^{-1}\, M_\Sun$.}
\vspace{0.5cm}
\label{f:fsat}
\end{figure}

The space densities, large-scale clustering strengths, and satellite
fractions of z2SFG descendants and various observed samples are
plotted in Figures \ref{f:nvsr0} and \ref{f:fsat}.  The uncertainties
on $r_0$ and $f_{\rm sat}$ for the z2SFG descendant halos reflect the
uncertainty in $\Mmin$ and the uncertain fate of satellites within
merged subhalos.  In contrast, the uncertainty in the number densities
of z2SFG descendant halos reflects the error in the observed z2SFG
number density, with the additional uncertainty due to mergers.  In
other words, the observed z2SFG number density is multiplied by the
fraction of z2SFG halos that survive to $z\sim1$ and $z\sim0$ in order
to deduce the number density of z2SFG descendants at these epochs. The
uncertainty in the halo number density due to the uncertainty in
$\Mmin$ thus is not included in the uncertainty in the observed z2SFG
descendants.  After discussing these constraints, we compare the
distribution of z2SFG descendant halo masses with the halo masses of
observed galaxies at lower redshifts, as inferred from halo occupation
modeling.  The most basic observational constraint, the abundance of
galaxies of various types at multiple epochs, is considered first.

\subsubsection{Number Density}
\label{s:nn}

In the no-merger scenario, the comoving number density of z2SFGs is
constant with time, by construction.  Including the possibility of
z2SFG mergers (as determined by the merging/disruption of their
associated subhalos), results in only mildly more evolution than the
no-merger case: by $z\sim1$ the number density of z2SFG descendant
halos is 76\% of its $z\sim2$ value and by $z\sim0$ it is $53$\% of
that at $z\sim2$.  Recall that our merger prescription associates a
z2SFG merger with the merging/disruption of its associated subhalo,
which is likely an over-estimate of the true merging because of
simulation resolution effects.  Nonetheless, if we use the fraction of
surviving halos as a proxy for the galaxy number density at lower
redshift, more than half of the z2SFGs exist as independent entities
at $z\sim0$.

Figures \ref{f:nvsr0} and \ref{f:fsat} plot along the $x$-axis the
number density of z2SFG descendants at $z\sim1$ and $z\sim0$.  Figure
\ref{f:nvsr0} also plots along the $x$-axis the abundances of observed
galaxies as a function of luminosity and color for absolute $B$-band
magnitude thresholds at $z\sim1$ (in half magnitude steps from $-19.5$
to $-20.5$ for the overall sample, and from $-19$ to $-21$ for the
color-defined samples) and for absolute $r$-band magnitude bins at
$z\sim0$ (in one magnitude intervals from $-19$ to $-22$) based on the
luminosity functions at these epochs \citep{Willmer06, Bell04,
  Blanton03c}.  For reference, at $z\sim1$ $M_B^\ast=-20.6$ and at
$z\sim0$ $M_r^\ast=-20.4$. At both epochs, blue and red galaxies are
separated according to the observed bimodal distribution of optical
colors \citep{Baldry04,Bell04}.  We have also included data at
$z\sim0$ for overall magnitude threshold samples in order to
demonstrate the small difference between using magnitude bins and
thresholds.\footnote{Magnitude bins, rather than thresholds, are used
  at $z\sim0$ because $r_0$ is only available in magnitude bins for
  the color-defined samples.}

The mild evolution of the z2SFG descendant halo number density already
places strong constraints on the possible progeny of z2SFGs.  It has
been proposed that red galaxies are the descendants of z2SFGs.
However, there are simply far too many z2SFG descendants compared to
the observed number density of $z\sim1$ red galaxies for this to be
the case, even when including rather faint ($M_B=-19.5$) red galaxies.
Since the faint-end slope of the red galaxy luminosity function at
$z\sim1$ is shallow \citep[$\alpha=-0.5$;][]{Willmer06}, the number
density of all red galaxies to a limiting magnitude of $M_B=-18.0$
($4.4 \times 10^{-3}\,h^{3}$ Mpc$^{-3}$) still does not equal the
abundance of z2SFG descendant halos.  The abundance of an overall
(i.e. no cut on color) sample of galaxies brighter than $M_B=-20.0$ at
$z\sim1$ is much more similar to the abundance of z2SFG descendant
halos.  At $z\sim0$ the conclusions are similar.  The observed sample
at $z\sim0$ with the number density closest to the z2SFG descendant
number density is the overall $-21<M_r<-20$ sample.

These conclusions, based solely on the abundances of halos and
galaxies, suggest that at most a fraction of z2SFGs have evolved onto
the red sequence by either $z\sim1$ or $z\sim0$ \citep[see also][who
reached similar conclusions]{Moustakas02, Gilli07}.  The z2SFGs also
do not appear to evolve exclusively into the most luminous ($>L^\ast$)
population of galaxies at either epoch, as the number density of such
galaxies is much lower than the descendant halos of z2SFGs.  Rather,
the abundances of descendant z2SFG halos is most similar to typical
$\sim L^\ast$ galaxies at both $z\sim1$ and $z\sim0$.

Previous work by \citet{Adelberger05} came to the conclusion that by
$z\sim1$ z2SFG descendants had largely evolved into red galaxies.
This conclusion was based in part on an earlier estimate of the
observed number density of red galaxies at $z\sim1$ \citep{Chen03},
which was a factor of several larger than current, more accurate
estimates \citep{Willmer06}.  The number density of red galaxies at
$z\sim0$ was similarly somewhat higher than the more recent data used
herein \citep{Zehavi05}.  The lower number density of observed z2SFGs
used in \citet{Adelberger05}, compared to what we adopt from more
recent data, also made it appear in that work that z2SFGs evolved into
red galaxies by $z\sim0$.  Our use of more accurate and up-to-date
number densities at all epochs is a significant reason why our results
are a departure from the conclusions reached by \citet{Adelberger05}.

\subsubsection{Clustering Strength}

Figure \ref{f:nvsr0} plots the large-scale clustering strength, $r_0$,
as a function of number density for both observed galaxy samples and
z2SFG descendant halos at $z\sim1$ ({\it top panel}) and $z\sim0$
({\it bottom panel}).  The data samples include both overall and
color-defined samples, for a range of luminosities described above.
The clustering data for these samples come from \citet{Coil06b} and
\citet{Coil07} for $z\sim1$ and from \citet{Zehavi05} for $z\sim0$.
For the halos, the power-law fit to the correlation function is
computed by solving simultaneously for $r_0$ and $\gamma$ over the
range $1<r<10 \, h^{-1}$ Mpc.  The uncertainty on $r_0$ for the halos
is dominated by the uncertainty on $\Mmin$ at $z\sim2$. 

From the figure it is clear that the clustering of z2SFG descendant
halos is comparable to that of the most luminous red galaxies, at both
epochs.  This point has been used as evidence that z2SFGs evolve into
red galaxies by $z\sim1$ \citep[e.g.][]{Adelberger05}.  However, as
discussed in the previous section, the constraints from the abundance
of z2SFG descendants strongly disfavors this scenario.  Moreover,
$r_0$ is sensitive to the number of satellite galaxies, and thus
should not be used as a constraint without also considering the
observed constraints on the satellite fraction.

\begin{figure}[t]
\plotone{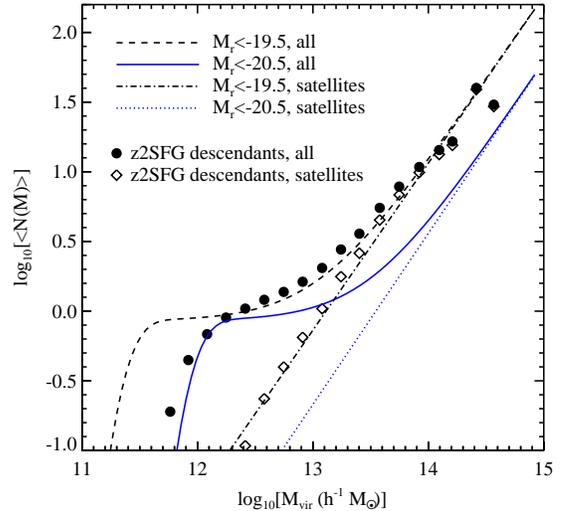}
\vspace{0.5cm}
\caption{The average number of z2SFG descendants per distinct halo at
  $z\sim0$ as a function of distinct halo mass for the overall
  descendant sample (\emph{filled circles}) and for the satellites
  only (\emph{diamonds}).  These results are compared to the average
  number of galaxies per halo as inferred from halo occupation
  modeling, for all galaxies with magnitudes $M_r<-20.5$ and
  $M_r<-19.5$, and for satellites only.}
\vspace{0.5cm}
\label{f:nm}
\end{figure}

\subsubsection{Satellite Fraction}
\label{s:fsat}

We now turn to a discussion of the satellite fractions of z2SFG halos
and their descendants, and of observed galaxies as inferred from halo
occupation modeling.  For z2SFG host dark matter halos, the satellite
(i.e. subhalo) fraction at $z\sim2$ is $11$\%, while at $z\sim1$ and
$z\sim0$ it is $24-39$\% and $31-59$\%, respectively. The lower and
upper bounds in satellite fraction depend primarily on the assumption
of merging or no merging of the satellites within destroyed subhalos.
Figure \ref{f:fsat} compares these satellite fractions to satellite
fractions derived for various data sets at $z\sim1$
\citep{Zheng07}\footnote{\citet{Zheng07} define a halo to be the mass
  enclosed within a region 200 times the mean density of the Universe
  (rather than the critical density as is used herein), and so their
  satellite fractions will be somewhat larger than those calculated
  using our halo definition.} and $z\sim0$.  The derived satellite
fractions were estimated by simultaneously fitting the observed number
densities and projected two-point correlation functions of the galaxy
samples with halo occupation models \citep[see e.g.][for
details]{Tinker05, Zheng07}.  The satellite fractions for the
color-defined samples at $z\sim0$ are reported here for the first
time.

From this plot it is clear that the z2SFG descendant halos have higher
satellite fractions than any overall luminosity-limited sample at
either $z\sim1$ or $z\sim0$ (except perhaps the faintest sample at
$z\sim0$).  Only the fainter ($-21<M_r<-19$) red galaxy samples at
$z\sim0$ have comparable satellite fractions.  On the other hand,
these faint red galaxies have large-scale clustering strengths that
are lower than the z2SFG descendant halos (Figure \ref{f:nvsr0}) and
they are thus not the likely descendants of z2SFGs.  The satellite
fraction of z2SFG descendants is higher than that of a typical
mass-selected sample of halos because, while the sample is
mass-selected at $z\sim2$, it is not mass-selected at lower redshifts
due to the different evolutionary histories of subhalos and distinct
halos (described in more detail in $\S$\ref{s:dsm}).

In order to better illustrate how the satellite fraction and
clustering strengths are related, we lower the satellite fractions of
the z2SFG descendant halos at $z\sim 1$ and $z\sim 0$ so that they
agree with the observed satellite fractions of $\sim L^\ast$ galaxies
at both epochs.  We do this by removing a fraction of the oldest
subhalos (i.e. those subhalos that accreted at the highest redshifts)
such that the resulting halo satellite fractions are 17\% at $z\sim1$
and 15\% at $z\sim0$.  In other words, at $z\sim0$ we have removed
60\% of available z2SFG descendant satellites.  As seen in Figure
\ref{f:nvsr0}, it is clear that the clustering of the z2SFG descendant
halos with lower satellite fractions is in much better agreement with
the clustering of observed overall $\sim L^\ast$ galaxies at both
epochs.  This result is not strongly affected by removing the oldest
subhalos as opposed to a random set; we remove the oldest because it
is these subhalos that are most likely to have faded significantly in
luminosity, and thus may plausibly not be included in a sample of
$\sim L^\ast$ galaxies at lower redshift.  Moreover, reducing the
satellite fraction to zero results in $r_0$ values of 4.2 and
4.9$h^{-1}$ Mpc at $z\sim1$ and $z\sim0$, respectively.  These numbers
bracket the observed values near $\sim L^\ast$. The implications of
this exercise are discussed more fully in the sections that follow.

\begin{figure}[t]
\plotone{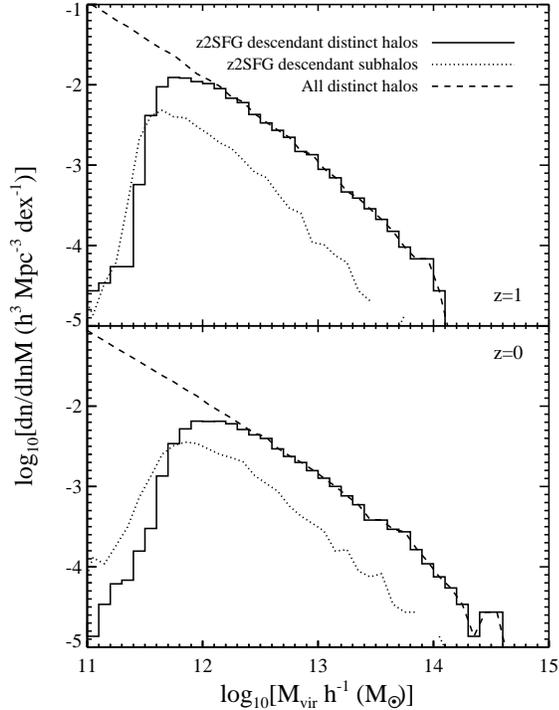}
\vspace{0.5cm}
\caption{Mass function for z2SFG distinct halo descendants
  (\emph{solid line}), z2SFG subhalo descendants (\emph{dotted line}),
  and all distinct halos ({\it dashed line}), at $z\sim1$ ({\it top
    panel}) and $z\sim0$ ({\it bottom panel}), not including subhalos.
  For subhalos, the mass refers to the mass it had at the epoch of
  accretion.  Note that for masses $\gtrsim10^{12} \,h^{-1}\,M_\Sun$
  the halos that contained z2SFGs constitute the vast majority of all
  halos above this mass at $z\sim1$ and $z\sim0$.}
\vspace{0.5cm}
\label{f:mf}
\end{figure}

\subsubsection{Insights from Halo Occupation Modeling}
\label{s:hod}

Halo occupation modeling has emerged as a powerful tool to analyze
observational data \citep[e.g.][]{Seljak00, Berlind02, Scoccimarro01,
  Bullock02, Kravtsov04, Zehavi05, Tinker05, Cooray06, Tinker07b,
  Zheng07, vdB07}.  The model quantifies the statistical distribution
of galaxies within dark matter halos, specifying the probability that
a halo of mass $M$ contains $N$ galaxies of a particular type. A given
halo occupation distribution maps nearly uniquely onto a two-point
correlation function, so the occupation function can be constrained by
clustering measurements.  In the context of this model, ``halos''
refers only to distinct halos, and not the subhalos within them.

The function $\langle N(M)\rangle$ specifies the mean number of
galaxies in a halo of mass $M$ and is typically split into two terms,
one describing the number of central galaxies, $\langle N_c\rangle$,
and the other describing satellite galaxies, $\langle N_s\rangle$ (the
dependence on halo mass is implicit in these functions).  When
constraining $\langle N(M)\rangle$, the approach taken is to assume
that $\langle N_c\rangle$ rises from zero to one at some mass scale,
with a rapidity constrained by the data.  The number of central
galaxies never rises above one by definition.  The number of satellite
galaxies is taken to be a power-law in halo mass, $\langle
N_s\rangle\propto M^\alpha$, where $\alpha$ is usually found to be
near unity.  This form is well-motivated both by results from
hydrodynamic \citep{Zheng05} and dissipationless simulations
\citep{Kravtsov04}, and observations \citep[e.g.][]{Lin04a}.  Details
of the halo occupation results presented in this section can be found
in \citet{Tinker05}, \citet{Tinker06}, and \citet{Tinker07b}, although
the fits presented herein are new because they utilize a halo mass
definition that is identical to the mass definition used throughout
the rest of this paper -- the mass enclosed within a region that is 200
times the critical density of the Universe.

Figure \ref{f:nm} plots the $\langle N(M)\rangle$ that best matches
the abundance and clustering of observed galaxies from the SDSS survey
with magnitudes $M_r<-20.5$ and $M_r<-19.5$ at $z\sim0$.  The figure
shows explicitly the contribution of satellites to the full $\langle
N(M)\rangle$. The contribution due to centrals can be inferred by
subtracting the satellite contribution from the total.  These results
are compared to the average number of z2SFG descendants per halo,
where the contribution due to z2SFG descendants that are satellites is
also included.

It is clear that $\langle N(M)\rangle$ for the z2SFG descendants that
are satellites by $z\sim0$ agrees very well with that of observed
satellite galaxies with $M_r<-19.5$.  Moreover, the fact that the
overall $\langle N(M)\rangle$ for z2SFG descendants drops rapidly to
zero at roughly the same mass scale as observed galaxies with
$M_r<-20.5$ indicates that the z2SFGs that are central galaxies by
$z\sim0$ correspond closely to galaxies with $M_r<-20.5$.  This last
fact follows because the region where $\langle N(M)\rangle\lesssim 1$
is dominated by central galaxies.  Scatter in the relation between
halo mass and UV luminosity at $z\sim2$, as discussed in
$\S$\ref{s:z2}, will produce a softer roll off in $\langle
N(M)\rangle$ at lower masses for the z2SFG descendants than what is
shown in Figure \ref{f:nm}.  The effect of scatter is thus to
associate a fraction of z2SFG descendants that are centrals with
galaxies fainter than $M_r=-20.5$.  Howevever, as discussed in
$\S$\ref{s:conn}, the scatter is not expected to be substantial
because of the observed UV-luminosity-dependent clustering observed at
$z\sim2$.

This comparison vividly demonstrates that z2SFG descendants that are
satellites by $z\sim0$ evolve into observed faint, $M_r<-19.5$,
galaxies, while those descendants that remain distinct halos
(i.e. central galaxies) evolve into more luminous, $M_r<-20.5$,
galaxies.  Echoing the conclusions drawn from previous sections, it is
clear from this figure that z2SFG descendants do not generically
evolve into a single class of galaxies at later epochs.  Rather, the
evolution of z2SFG descendants that become satellites is qualitatively
different from those that become central galaxies.

The ratio between $\langle N_s\rangle$ for galaxies brighter than
$M_r=-20.5$ and $M_r=-19.5$ is $\sim30$\%, indicating that
approximately 30\% of galaxies brighter than $M_r=-19.5$ are also
brighter than $M_r=-20.5$.  In $\S$\ref{s:fsat} we found that keeping
only 40\% of the total number of z2SFG descendants that are satellites
yielded a sample of z2SFG descendants that had both clustering
strengths, satellite fractions, and abundances in good agreement with
the observed sample of galaxies at $z\sim0$ with $M_r\lesssim-20$.
The fact that these fractions are similar suggests a physical
motivation for reducing the satellite fraction of the z2SFG
descendants in order to produce better agreement with an observed
$z\sim0$ sample defined with respect to a luminosity cut.  A large
fraction of z2SFG descendants that become satellites are faint, and so
the overall sample of z2SFG descendants, which includes these faint
satellites, will not compare to any observed sample of galaxies
defined according to a simple luminosity cut.

\subsubsection{Descendant Halo Masses}
\label{s:dsm}

Finally, we turn to a discussion of the mass distribution of halos
hosting z2SFG descendants at lower redshifts.  Figure \ref{f:mf} plots
the mass function of distinct z2SFG descendant halos at $z\sim1$ and
$z\sim0$ and compares to the full mass function of distinct halos from
the simulation.  It is clear that z2SFG descendant halos with mass
$\gtrsim10^{12} \, h^{-1} M_\Sun$ constitute the vast majority of all
halos in that mass range at $z\sim1$ and $z\sim0$.  In other words,
virtually all halos today with mass $\gtrsim10^{12} \, h^{-1} M_\Sun$
contained at least one z2SFG.  In light of this fact, it thus appears
that the descendants of z2SFGs are not only the galaxies at the
centers of massive halos but rather constitute a wide variety of
objects in a variety of environments.  In fact, by far the most common
place to find the descendants of z2SFGs at $z\sim0$ is in $\sim
10^{12} \, h^{-1} M_\Sun$ halos.

It is furthermore clear from the figure that the average mass of
distinct z2SFG descendant halos is growing modestly with time.  At
$z\sim2$ the average distinct halo mass is $10^{11.8}\, h^{-1} \,
M_\Sun$, at $z\sim1$ it has increased to $10^{12.1}\, h^{-1} \,
M_\Sun$ and by $z\sim0$ it is $10^{12.3}\, h^{-1} \, M_\Sun$.  Note
that these average masses are not weighted by the number of
satellites/subhalos within each distinct halo.

Figure \ref{f:mf} also includes the mass function of z2SFG descendants
that are subhalos by $z\sim1$ and $z\sim0$.  Here the subhalo mass is
measured at the epoch of accretion.  Aside from the obvious fact that
the subhalos constitute only a fraction ($f_{\rm sat}$) of the total
halo population, the mass function of these subhalos clearly peaks at
a lower mass than for the distinct descendant halos.  This is not
surprising given the fact, for example, that the average redshift of
accretion for subhalos identified at $z\sim0$ is $z\approx0.5$.  The
mass that these subhalos had at the epoch of accretion is on average
only 55\% of the average mass of $z\sim0$ z2SFG descendant halos that
are not subhalos.  Thus, in the $\sim3.5\, h^{-1}$ Gyr since
$z\approx0.5$, z2SFG descendants residing at the centers of distinct
halos continued to grow, both in dark matter, gas, and presumably in
stars (via star formation), while those in subhalos did not grow in
dark matter or gas, and were likely subject to one or more
gas-starvation processes, thereby further halting the growth of the
galaxy embedded within the subhalo.  This difference between z2SFG
descendants that are subhalos and those that are distinct halos has
significant implications for the galaxies likely embedded within them.

\subsubsection{Sensitivity to Cosmological Parameters}

As discussed in $\S$\ref{s:z2}, we have made use of an $N$-body
simulation run with cosmological parameters advocated by the
\emph{WMAP3} results \citep{Spergel07} in order to assess the
sensitivity of our results to these new parameters.  In $\S$\ref{s:z2}
we found that the halos in the \emph{WMAP3} cosmology with clustering
most similar to observed z2SFGs have $\Mmin$ lower by $0.3-0.4$ dex
compared to the \emph{WMAP1} cosmology.  However, once we have
identified a population of halos with comparable clustering to the
observed population in both cosmologies, their subsequent evolution to
lower redshift is almost identical.  In particular, we have followed
the evolution of distinct halos (not subhalos) in the \emph{WMAP3}
cosmology and find that $r_0$ of the descendants is $5-10$\% lower
compared to the \emph{WMAP1} cosmology at both $z\sim1$ and $z\sim0$.
Furthermore, the fraction of halos that merged away by $z\sim1$ and
$z\sim0$ differs by less than 5\% for the two cosmologies.  Note that
since we use the observed number density of z2SFGs multiplied by the
fraction of halos that survive to later epochs, rather than the number
density of halos themselves, our quoted number densities of
descendants are insensitive to cosmological parameters (see
$\S$\ref{s:z10} for more details).  The differences between these two
cosmologies thus has a negligible impact on our conclusions regarding
the descendants of z2SFGs.

\section{Discussion}
\label{s:disc}

In the preceding section we found that the observed abundance and
clustering of z2SFGs implied that every halo at $z\sim2$ above a mass
of $10^{11.4}\, h^{-1} \, M_\Sun$ contains one z2SFG.  By $z\sim0$,
the overall population of halos that once hosted z2SFGs, and survived
to $z\sim0$, has a comoving number density comparable to that of $\sim
L^\ast$ galaxies ($M_r\lesssim-20$; while $M_r^\ast=-20.4$), a
clustering strength comparable to luminous ($M_r\lesssim-21$) red
galaxies, and a satellite fraction comparable to $\sim L^\ast$ or
fainter red galaxies.  We found that a subsample of z2SFG descendant
halos where 60\% of the satellites were removed resulted in much
better agreement with the clustering strength, satellite fraction, and
number density of $\sim L^\ast$ galaxies at $z\sim0$.  Comparison to
halo occupation modeling of observed galaxy samples revealed that the
z2SFG descendants that are central galaxies most closely resemble
observed galaxies brighter than $M_r\sim-20.5$ while the satellite
properties compare more favorably to galaxies brighter than
$M_r\sim-19.5$.  The average redshift at which z2SFG descendant
satellites at $z\sim 0$ were accreted onto their parent halo was
$z\approx0.5$, and the mass of the halo associated with these
satellites when they were accreted was only 55\% of the mass of
distinct halos at $z\sim0$, on average.  This indicates that the
growth of subhalos (and the galaxies within them) was significantly
less than that of distinct halos.

These results have straightforward implications regarding the
connection between z2SFGs and their descendants at $z\sim0$.  First,
it is abundantly clear that z2SFGs do not evolve into any single
sample of galaxies defined either according to a luminosity cut or a
luminosity and color cut.  In particular, they do not evolve
exclusively into massive red galaxies, either by $z\sim1$ or $z\sim0$.
Rather, the subsequent evolution of z2SFGs is more nuanced.  Merger
trees extracted from simulations indicate that the number density of
z2SFG descendants at $z\sim0$ is $\sim50$\% of the number density of
z2SFGs because some descendants merge together between $z\sim2$ and
$z\sim0$.  This is an upper bound on the evolution of the number
density because halos in our simulations can merge or disrupt without
the subsequent merger or disruption of the galaxy within it due to
simulation resolution effects or possible baryonic effects.  Of the
remaining descendants, $\sim70$\% are centrals and $\sim30$\% are
satellites.  The central galaxies have properties similar to observed
galaxies with $M_r\lesssim-20.5$.  In particular, they reside
predominantly in halos with mass $M\gtrsim10^{12}\, h^{-1} M_\Sun$.

An anecdotal consequence is that our own Galaxy, with halo mass $\sim
10^{12} \, h^{-1} \,M_\Sun$ \citep{Klypin02}, was once likely a z2SFG.
This conclusion is in accord with various properties of the bulge of
our Galaxy.  For example, the bulge has a stellar mass of $\sim
10^{10}\,M_\Sun$ that is thought to have formed over a short period of
time ($<1$ Gyr) roughly 10 Gyr ago \citep{Zoccali03, Ferreras03}.
These facts imply that the star-formation rate during the formation of
the bulge was $>10\, M_\Sun$ yr$^{-1}$.  The epoch of formation of the
bulge and its high star-formation rate imply that it would have likely
been detected as a z2SFG \citep[see also][who reached similar
conclusions]{Pettini06}.

In contrast, the z2SFG descendants that become satellites by $z\sim0$
are much more similar to a fainter sample of observed galaxies
($M_r\lesssim-19.5$).  This is not surprising in light of the
different histories of satellite compared to central galaxies as
inferred from the histories of their dark matter halos.  Since
satellites fell into their parent halo on average at $z\approx0.5$
($3.5\,h^{-1}$ Gyr ago), their growth was significantly retarded
relative to central galaxies.  Satellites grew less not only because
they could not accrete new gas but also because the gas they possessed
at infall was likely prohibited from cooling to form stars (or removed
altogether) due to one or more quenching processes, such as gas
strangulation, harassment, or ram-pressure stripping.

Not all satellites are required to be so faint; our results indicate
that $\sim40$\% are brighter than $M_r\sim-20.5$.  It is plausible
that the $\sim60$\% required to have $-19.5\lesssim M_r\lesssim-20.5$
were the satellites accreted at the earliest epochs, because such
satellites would have the least amount of growth between $z\sim2$ and
$z\sim0$.  Simulated star-formation histories coupled with a stellar
population synthesis code \citep{Bruzual03} confirm that galaxies with
properties similar to z2SFGs can easily evolve into galaxies with
$M_r\sim-19.5$ if their star-formation is truncated at $z\sim0.5-1$.
For example, a galaxy at $z\sim2$ with stellar mass $10^{10}\,h^{-2}\,
M_\Sun$ and a star-formation rate of 10 $M_\Sun$ yr$^{-1}$ that
steadily declines to zero at $z\sim1$ fades from $M_r=-21.5$ to
$M_r=-19.2$ by $z\sim0$.

Similar conclusions hold for the descendant halos at $z\sim1$.  In
particular, from analysis of the Millennium simulation merger trees we
find that $\sim25$\% of z2SFG descendants have merged away by this
time.  Of the remaining halos, $\sim25$\% are satellites and
$\sim75$\% are central galaxies.  If half of the satellites are
fainter than $\sim L^\ast$, then the remaining descendant halos that
are satellites and centrals have a number density, clustering
strength, and satellite fraction comparable to observed $\sim L^\ast$
galaxies at $z\sim1$.

There has been some debate in the literature over whether or not the
optical color selection technique used to select z2SFGs misses a
significant population of redder (i.e. older and/or dustier) galaxies
\citep[e.g.][]{Franx03, Daddi04, vanDokkum06, Smail02}.  Under our
assumption of a tight correlation between rest-frame UV luminosity and
halo mass, we have demonstrated that at $z\sim2$ roughly every halo
with mass above $10^{11.4} \,h^{-1}\, M_\Sun$ contains one z2SFG.  It
would be rather surprising if each halo also contained another
comparably massive galaxy not detectable with the optical color
selection technique, if only by analogy with lower redshift, where the
vast majority of $\sim L^\ast$ galaxies live alone in their dark
matter halos \citep[e.g.][]{Zheng07}.  Thus, while there may well be a
population of massive ($M_\ast\gtrsim 10^{11} \,h^{-2}\,M_\Sun$), red
galaxies with a correspondingly low space density ($\sim10^{-4}\, h^3$
Mpc$^{-3}$), that is missed with the optical color selection technique
\citep{Franx03, Daddi04, vanDokkum06}, it is hard to imagine that this
could be the case at lower galaxy masses.  Since our conclusions are
most sensitive to lower mass galaxies, they are not influenced by this
possible distinct population of more massive, less numerous, red
galaxies.

There are two basic facts about dark matter halos and their evolution
that, when combined, provide both a qualitative understanding of our
results and more general insights into the evolution of populations of
galaxies across time.  The first fact is that, to with a factor of
$\sim2$, the space density of a set of halos above a mass threshold
does not change from high redshift to the present.  For example, of
the halos with $M\geq10^{11.4}\, h^{-1}\, M_\Sun$ at $z\sim2$, 50\%
survive to $z\sim0$.  For halos more massive than $10^{11.9}\,
h^{-1}\, M_\Sun$ and $10^{12.4}\, h^{-1}\, M_\Sun$ at $z\sim2$, the
fraction that survive to $z\sim0$ increases to 60\% and 70\%,
respectively.  The second fact is that, by and large, the ranking of
halos by mass is preserved as the Universe evolves.  For example, the
most massive halos at $z\sim2$ evolve to the most massive halos today.
If, as we have argued, the z2SFGs are an at least approximately halo
mass selected sample, then these two basic facts make less surprising
the conclusion that z2SFGs evolve largely into a roughly halo mass
selected sample of abundant galaxies at $z\sim0$.  As described above,
it is primarily the uncertain evolution of satellite galaxies that
complicates this simple description.

These facts can also shed light on the likely descendants of the
massive red galaxies discussed above.  If such galaxies are in the
most massive halos \citep[as suggested by their high stellar mass and
clustering strength;][]{Quadri07}, then it is these galaxies that will
evolve, largely intact, to the centers of rich groups and clusters by
$z\sim0$.  Better constraints on the clustering and abundances of
these observed massive red galaxies are required to make more detailed
statements.

Finally, these basic facts can connect our discussion herein to the
fates of star-forming galaxies at $z>2$.  Indeed, much of the previous
high-redshift modeling efforts discussed in the Introduction focused
on star-forming galaxies at $z\sim3$, the so-called Lyman-Break
Galaxies (LBGs).  If we apply our methods to match the observed
clustering of LBGs \citep[$r_0=4.0\pm0.6\,h^{-1}$
Mpc;][]{Adelberger05} with a set of halos in the Millennium simulation
above some minimum mass, we find a best-fit $\Mmin =
10^{11.1\pm0.2}\,h^{-1}\,M_\Sun$.  Analysis of the z2SFG halo merger
trees indicates that every z2SFG halo contains one or more of these
LBG halos in their history and approximately 70\% of LBGs evolve into
z2SFGs \citep[in agreement with earlier work; e.g.][]{Adelberger05}.
While the link between LBGs and z2SFGs appears strong, there are
important subtleties one must keep in mind when comparing LBGs to
z2SFGs.  In particular, the clustering of these two populations has
been measured for all objects above a common apparent magnitude limit
($\mathcal{R}=25.5$), and thus the LBGs, being more distant, are an
intrinsically more luminous sample compared to the z2SFGs.  However,
the luminosity functions of these two classes of galaxies are nearly
the same \citep{Reddy07}, suggesting that if the two samples were
defined with respect to the same absolute luminosity limit, their
evolutionary bond would be even stronger.  Nonetheless, the clear link
between LBGs and z2SFGs implies that LBGs share the same varied fates
as those we have outlined for z2SFGs.

The most significant uncertainty in our analysis is the clustering
strength of z2SFGs and the subsequent merging of their descendants.  A
more quantitative understanding of the fate of these z2SFGs will thus
require a more accurate accounting of their clustering strengths, on
the observational side, and a more sophisticated treatment of z2SFG
mergers on the theoretical side.  In addition, clustering measurements
on smaller scales ($\lesssim1$Mpc) and over a range of rest-frame
luminosities will allow a more detailed modeling of the connection
between z2SFGs and halos at $z\sim2$ than has been presented here and
will provide a critical test of our assumption of a tight correlation
between rest-frame UV luminosity and halo mass at $z\sim2$.

With more accurate data over the interval $0\lesssim z\lesssim 2$, the
goal will ultimately be to model the galaxy-halo connection as a
function of galaxy properties such as color and luminosity, or
star-formation rate and stellar mass, in narrow redshift intervals.
The galaxy-halo connections at higher redshift can then be evolved
into the connections at lower redshift with the aid of halo merger
trees.  The merger trees are the key to this approach, as they put the
static (fixed redshift) halo occupation models into motion.  Such an
approach has only recently become feasible, due both to the vastly
increased data at these epochs and to a converging set of cosmological
parameters that removes significant uncertainties in the properties
and histories of dark matter halos \citep[see e.g.][]{Zheng07,
  MWhite07, Conroy07b}.  A simplified version of this approach has
been explored in the present work.  A more detailed version will be
pursued elsewhere.

\acknowledgments

We thank Alison Coil, Eric Hayashi, Idit Zehavi, and Zheng Zheng for
providing and assisting in the interpretation of their data and
results. We also thank J\'er\'emy Blaizot, Alison Coil, Gabriella De
Lucia, Andrey Kravtsov, Jerry Ostriker, Max Pettini, Ryan Quadri, Risa
Wechsler, Martin White, and Simon White for numerous helpful comments
on an earlier draft.  AES thanks Guinevere Kauffmann, J\'er\'emy
Blaizot and Gabriella De Lucia for their generous hospitality at the
Max Planck Institute for Astrophysics, and acknowledges support from
the David and Lucile Packard Foundation and the Alfred P. Sloan
Foundation.  GL works for the German Astrophysical Virtual Observatory
(GAVO), which is supported by a grant from the German FederalMinistry
of Education and Research (BMBF) under contract 05 AC6VHA.  The
Millennium Simulation databases used in this paper and the web
application providing online access to them were constructed as part
of the activities of the German Astrophysical Virtual Observatory.
This work made extensive use of the NASA Astrophysics Data System and
of the {\tt astro-ph} preprint archive at {\tt arXiv.org}.

\end{document}